\documentclass[12pt]{revtex4}
\usepackage{graphicx,epsfig}
\usepackage{amsmath}
\usepackage{bm}
\usepackage{color}
\usepackage{colordvi}
\usepackage{citesort}

\begin{document}

\title[Time-resolved relaxation of  X-ray excited gallium arsenide]{Time-resolved observation of band-gap shrinking and electron-lattice thermalization within X-ray excited gallium arsenide}

\author{Beata Ziaja$^{1,2}$\footnote{Corresponding authors: B. Ziaja  (ziaja@mail.desy.de) and N. Medvedev (nikita.medvedev@desy.de).},
Nikita Medvedev$^{1}$, Victor Tkachenko$^{1}$, Theophilos Maltezopoulos$^{3}$, and  
Wilfried Wurth$^{1, 3, 4}$}

\address{$^{1}$Center for Free-Electron Laser Science CFEL, Deutsches Elektronen-Synchrotron DESY, Notkestrasse 85,  22607 Hamburg, Germany\\
$^{2}$Institute of Nuclear Physics, Polish Academy of Sciences, Radzikowskiego 152, 
31-342 Krak\'ow, Poland\\
$^{3}$ Department of Physics and Center for Free-Electron Laser Science CFEL,  University of Hamburg, Luruper Chaussee 149, 22761 Hamburg, Germany\\
$^{4}$ DESY Photon Science, Deutsches Elektronen-Synchrotron DESY, Notkestrasse 85, 22607 Hamburg, Germany}

\begin{abstract}
Femtosecond X-ray irradiation of  solids excites energetic photoelectrons that thermalize on a timescale of a few hundred femtoseconds. The thermalized electrons exchange energy with the lattice and heat it up. Experiments with X-ray free-electron lasers have unveiled so far the details of  the electronic thermalization. In this work we show that the data on transient optical reflectivity measured in GaAs  irradiated with femtosecond  X-ray pulses can be used to follow electron-lattice relaxation up to a few tens of picoseconds. With a dedicated theoretical framework,  we explain the so far unexplained reflectivity overshooting as  a result of  band-gap shrinking. We also obtain predictions for a timescale of electron-lattice thermalization, initiated by conduction band electrons in the temperature regime of a few eVs. The conduction and valence band carriers were then strongly non-isothermal.  The presented scheme is of general applicability and can stimulate further studies of relaxation within X-ray excited narrow band-gap semiconductors.
\end{abstract}

\date{\today}
 
\keywords{Free-electron lasers; transient optical properties; electron-lattice thermalization time}

\maketitle

\section*{Introduction}

Intense ultrashort radiation available at the free-electron-laser (FEL) facilities all over the world \cite{LCLS,SACLA2,FLASH,EXFEL,elletra} enables structural studies of matter with atomic resolution \cite{henry1,henry2},  and,  with a pump-probe scheme,  investigation of transient states of matter with a few femtosecond resolution. A pump pulse (FEL, optical, THz) initiates a transition which is probed with another laser pulse delayed by a certain time with respect to the pump pulse. It was already found during the first solid-state experiments with FELs that an intense FEL laser pulse excites many electrons within the irradiated solid.  This leads to a transient change of the optical properties of the target which follows the evolution of the electron density. In particular,  this allows to estimate the timescale of the delay between pump and probe pulse.  Therefore, in the first \cite{gahl,alexander2} and later experiments \cite{diagnostic1,diagnostic2,diagnostic3,harmand,riedel} measuring transient optical properties of solids the detected changes have been used to design non-destructive time-delay diagnostic and pulse characterization tools for FELs. 

However, this experimental scheme also opens possibilities to extract information about the  physical processes following the relaxation of  the laser-excited materials. Similar techniques have been used in two-color experiments with optical lasers to extract information on the electron-phonon coupling coefficient \cite{hostetler} or to measure  the temperature-dependent electron-lattice thermalization time \cite{delfatti, delfatti1,cho}.  Transient changes of  transmission within X-ray irradiated GaAs  were reported in Ref.\ \cite{durbin} and interpreted as a signature of non-equilibrium relaxation effects within the irradiated solid.  

In this study we revisit the results of  the experiments measuring the transient optical reflectivity in GaAs irradiated  with femtosecond pulses of soft and hard X-ray radiation \cite{gahl,krupin}. GaAs is a direct band-gap semiconductor. Its band-gap width of $E_{gap}=1.42$ eV at room temperature \cite{ioffe} is slightly smaller than the energy of the optical photons at $800$ nm ($E_{\gamma}=1.55$ eV). Thus, in addition to the free carrier absorption, optical photons may also trigger electron excitation from the valence to the conduction band, which then becomes the dominant channel of the energy absorption \cite{sturge,ioffe}. The rate of this interband excitation depends on the band occupations and on the band-gap width. As the band-gap width in semiconductors is a function of the lattice temperature, the interband transition rate can reflect  the transient changes of the lattice temperature. 

Electronic excitation and relaxation processes following the FEL irradiation of solids are well understood \cite{ziaja, medvedev_NJP_2010,fitting}. The arriving FEL photons excite electrons from the valence to the conduction-band and, at sufficiently high photon energies, also from the atomic inner shells. In light elements the inner-shell electron excitation is followed by an Auger decay, resulting in the emission of another electron into the conduction-band. Energetic electrons within the conduction band can excite more electrons from the valence band through the electron impact ionization. The electrons  also interact with each other, exchanging energy. This leads to a fast thermalization of the conduction band electrons. Its timescale $\tau_{el}$  depends on the photon energy and fluence. At a typical few tens of femtoseconds FEL pulse durations, the electron thermalization timescales were reported to be $\leq 200$ fs both by theory \cite{ziaja, medvedev_NJP_2010} and experiment \cite{gahl,krupin}. Similar timescales were found in experiments with optical excitation \cite{knox,oudar,kim}. A recent study \cite{medvedev_CPP} showed that the electron thermalization following FEL pulses  is mainly determined by the secondary electron cascading, which timescale increases with the increasing photon energy. 

However,  thermalization between the valence and conduction band may take longer time than the carrier thermalization within the conduction band. The  relation between these two timescales depends on the ratio of the effective masses of holes and electrons. In GaAs holes are almost $10$ times heavier than the conduction band electrons. This allows to expect that the absorbed energy (brought by X-ray photons) will be initially shared among the light conduction-band electrons and then slowly transferred to the heavy holes. Conduction and valence-band carriers  then remain for some time strongly non-isothermal. Similar carrier non-isothermality has been observed in \cite{joshi, tomassi} and recently in \cite{bismuth}. 

\begin{figure*}
\includegraphics[width=12cm]{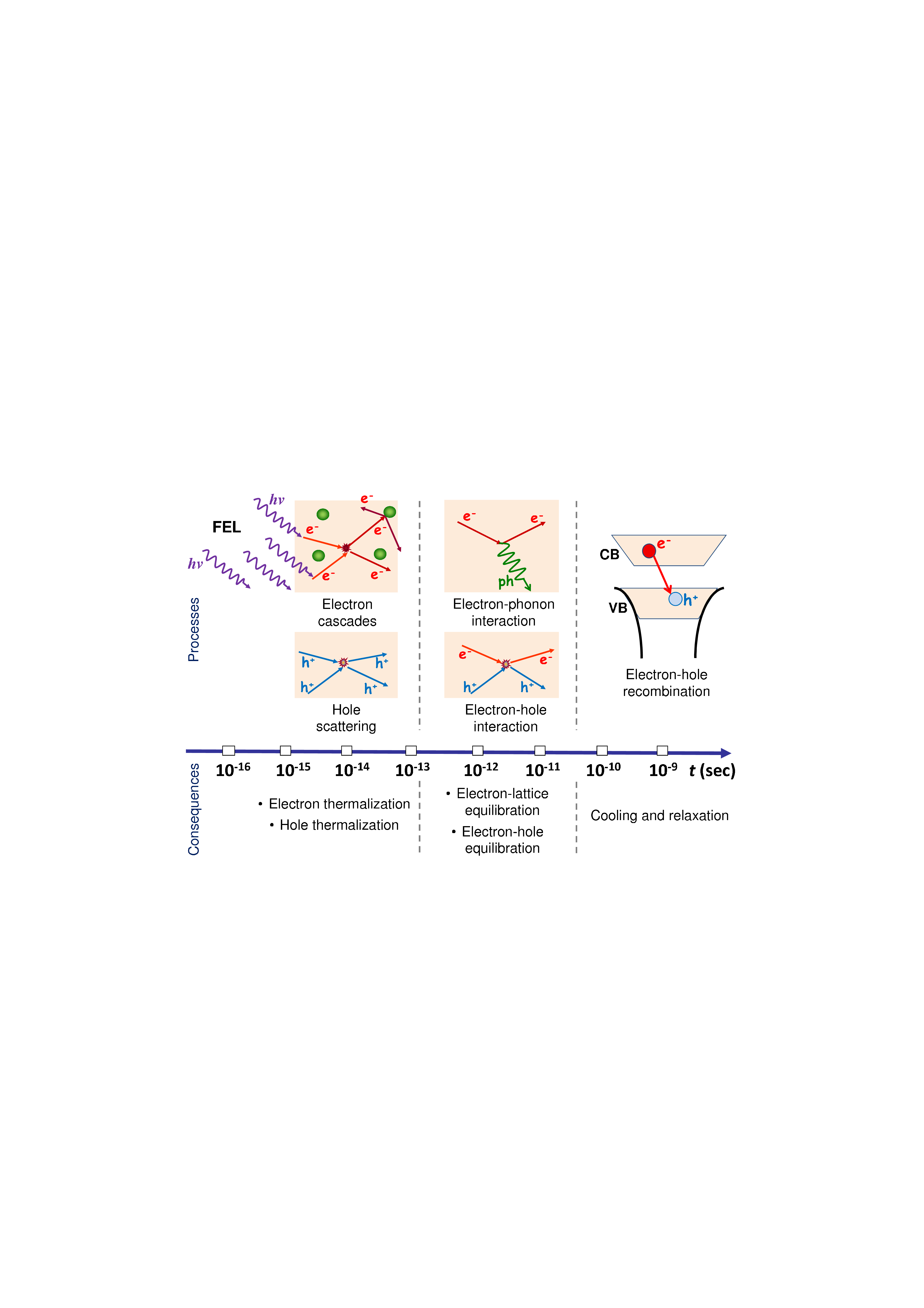}

\caption{(Color online) Timescales for predominant excitation and relaxation processes in X-ray irradiated GaAs.} 
\label{scales}
\end{figure*}

Already after the appearance of  first photoelectrons within the conduction band, the energy exchange between free carriers and lattice starts. In semiconductors the timescale of this process $\tau_{el-latt}$ is of the order of picoseconds, which is typically much longer than the electron thermalization time \cite{delfatti,delfatti1}. While the temperature of the lattice increases, the band gap shrinks. Based on the experimental data, phenomenological fits were constructed to describe the band-gap shrinking as a function of the  lattice temperature \cite{ioffe,dresselhaus,blakemore,beye_NJP}. We consider here the case when the densities of excited carriers  are low and do not lead to an additional band-gap shrinking \cite{spataru}. After the free carriers thermalize, recombination processes begin to contribute. In GaAs both radiative and non-radiative recombination occur \cite{saleh}. However, the typical recombination times for GaAs are of the order of  $\tau_{rec}\sim100$ ns \cite{saleh}, which is much longer than the electron-lattice thermalization time \cite{delfatti}. Fig.\ \ref{scales} shows the timescales of the predominant excitation and relaxation processes in X-ray irradiated GaAs.

In this work we show that the experimental data on the transient reflectivity allow to  follow in time the relaxation of  FEL excited GaAs on timescales up to a few tens of picoseconds. We follow the ideas from the optical measurements described in Refs. \cite{hostetler, delfatti,delfatti1}, however, the photon wavelengths considered in this work are in X-ray regime. They heat the solid to much higher temperatures ($\sim$ few eVs) than those considered in \cite{delfatti,delfatti1}. This excitation results in the characteristic 'reflectivity overshooting',  i.e., after the relaxation the reflectivity reaches an equilibrium value higher than its initial value before the excitation.  This effect has not been explained so far  \cite{durbin1}.  

Our proposed theoretical framework uses rate equations to describe the evolution of electron distribution as a function of time. The rate equations are coupled with the two-temperature model \cite{anisimov}, describing the electron-lattice equilibration. The Drude model is applied to follow the  transient reflectivity as a function of free-carrier density. Here,  this model is extended beyond the free-carrier absorption framework \cite{dresselhaus} and it also includes the predominant contribution from interband transitions. Comparing this model to the available data, we identify the reflectivity overshooting as a result of the band-gap shrinkage and obtain predictions for the electron-lattice thermalization time as a function of pulse fluence and photon energy.  If more experimental data were available, the presented scheme could be applied  for a quantitative study  of excitation and relaxation times in X-ray excited GaAs and in other narrow-band-gap semiconductors for which $\tau_{el} << \tau_{el-latt} << \tau_{rec}$. 

\section*{Results}

In experiments with X-ray FELs \cite{krupin,gahl} relative changes of transient reflectivity within GaAs have been measured, $\Delta R/R \equiv (R-R_0)/R_0$, as a function of  the time delay between the FEL pump pulse and the optical probe pulse. In the experiment with soft X-rays by Gahl et al. \cite{gahl},            
FEL photons had the energy of  $40$ eV. In the experiment by Krupin et al. \cite{krupin}, photons of $800$ eV energy were used. Various FEL fluences have been applied for pumping the material. The FEL pulse duration did not exceed a few tens of fs. Optical probe pulses were of low intensity so that they did not cause any damage in GaAs. Their wavelengths were: (i)  $800$ nm ($\sim1.55$ eV) in \cite{krupin} and (ii)  both $800$ nm and  $400$ nm ($\sim3.1$ eV) in \cite{gahl}. The error of fluence estimation in the earlier experiment \cite{gahl} performed in 2008 is expected to be much larger than in \cite{krupin} performed in 2012 as the pulse diagnostics methods had been significantly improved.

Both experiments observed a characteristic shape of transient reflectivity curve, $\Delta R/R$. On the timescale of  a few hundreds of  femtoseconds since FEL irradiation, one can see an initial ultrafast drop and rise of optical reflectivity (Figs. 2-3 in Ref.\ \cite{gahl}, Fig. 5 in  Ref.\ \cite{krupin}). These changes are followed by a slower relaxation of reflectivity towards an equilibrium value on a timescale of tens of ps. The transient reflectivity then overshoots the initial value. This is in contrast to the behaviour of irradiated insulators (e.g., silicon dioxide \cite{harmand} and silicon nitride \cite{krupin}). In what follows we prove that this effect can occur only for semiconductors with a band gap slightly smaller than the energy of the probing optical photon. In this case, the lattice temperature-sensitive electronic transitions between valence and conduction band triggered by the optical photons  influence significantly the transient optical properties of the material. As the excited carrier densities (maximally of the order of a few permilles of the initial valence electron density) are much too low to affect the band gap \cite{spataru}, the transient optical properties then only reflect the band-gap shrinking due to the increasing temperature of the lattice. 

With our modeling tool we performed simulations of the transient reflectivity changes within GaAs for the experimental conditions as described in Refs. \cite{krupin, gahl}. Our tool can describe optical properties within GaAs bulk, far from its surface. Therefore, all emitted electrons are assumed to stay within the material, i.e., they are emitted to the conduction band (below the continuum level of the material). The charge neutrality is then preserved.

First, we estimated the final electron-hole density after FEL irradiation, knowing the pulse fluence and  the photoabsorption cross section at a given photon energy. We then followed the increase of the electron-hole density until the maximal density was reached.  Such behaviour is typical for FEL irradiated semiconductors as described, e.g., in \cite{ziaja, medvedev_NJP_2010}.

After the maximal electron-hole density is reached, the system starts to relax. Electron-lattice thermalization and electron-hole recombination are the predominant relaxation channels. In GaAs the latter one contributes less significantly, as the typical recombination timescales are of the order of  $100$ ns \cite{saleh}. Electron-lattice thermalization in GaAs is expected to act on ps time scales  \cite{delfatti, delfatti1}. The resulting lattice heating leads to a shrinkage of  the band gap \cite{blakemore}, when compared with its initial width of   $E_{gap}=1.42$ eV at 300 K.  In what follows we demonstrate that this affects the interband photoabsorption, which results in the experimentally observed 'overshooting' of the transient reflectivity. 

The extended Drude model including the interband contribution is used to calculate the transient reflectivity change from the dielectric function, $\epsilon$, which is the function of optical coefficients, $n$ and $k$. See Supplemental Material at [URL will be inserted by publisher] for more details on the model. The dielectric function is parametrized as in Ref.\ \cite{dresselhaus}:
\begin{equation}
\epsilon \equiv (n+i\,k)^2=\epsilon_{core}
-\sum_{j=e,h}{\omega_{p,j}^2 \over \omega_{\gamma}^2}\,
{ {(\omega_{\gamma}\tau_j)}^2 \over {{(\omega_{\gamma}\tau_j)}^2 + 1}  }
+\sum_{j=e,h} i \,{\omega_{p,j}^2 \over \omega_{\gamma}^2}\,
{ {\omega_{\gamma}\tau_j} \over {{(\omega_{\gamma}\tau_j)}^2 + 1}  },
\end{equation}
where $\epsilon_{core}$ describes all contributions to the dielectric function beyond the free-carrier absorption. Here,  $\epsilon_{core} =(n_{core}+i\,k_{interband})^2$, where $k_{interband}=\alpha\, \lambda_{\gamma}/4\pi$ describes the contribution from the transition between valence and conduction bands, using the interband absorption coefficient for a direct interband transition, $\alpha$,  parametrized as in Ref.\ \cite{dresselhaus} (Eq.\ (5.31) therein). The absorption coefficient scales with the photon energy   
$E_{phot}$, as $\alpha \sim \sqrt{E_{phot} -E_{gap}}/E_{phot} $. Band-gap shrinking is described with the phenomenological relation from \cite{blakemore}. The interband absorption coefficient also contains the matrix element, $<v|p|c>$, which couples states with the same electron wave vector in the valence and conduction bands. We parametrize it, using the measured absorption coefficients for GaAs from Ref. \ \cite{sturge} (Fig. 3  therein). The time $\tau_{e(h)}$ is the electron (hole) collision time and the frequency $\omega_{p,e(h)}$ is the plasma frequency for electrons (holes).  Carrier density $n_{e-h}$ determines  the plasma frequency of the material  $\omega_{p(e,h)} \sim \sqrt{n_{e-h}}$. The frequency $\omega_{\gamma}$ is the photon frequency and $\lambda_{\gamma}$ is its corresponding wavelength.

The average electron collision time $\tau_e$ is fitted in order to match the minimum of $\Delta R/R$ curve. The accuracy of the fit depends on the accuracy with which we can estimate the minimum reflectivity. This is determined by the time resolution of the data. An improvement of the experimental temporal resolution would improve the accuracy of all model fits performed.
The average hole collision time $\tau_h$ can be estimated from the electron one, using mass scaling relation for electron and hole collision frequencies \cite{ashcroft}. The initial value of the lattice temperature is $300$ K. Lattice temperature does not change much during the first $100 -200$ fs after the FEL irradiation (during electron thermalization) and so the band-gap width does not change either.   

\begin{figure*}
\includegraphics[width=15cm]{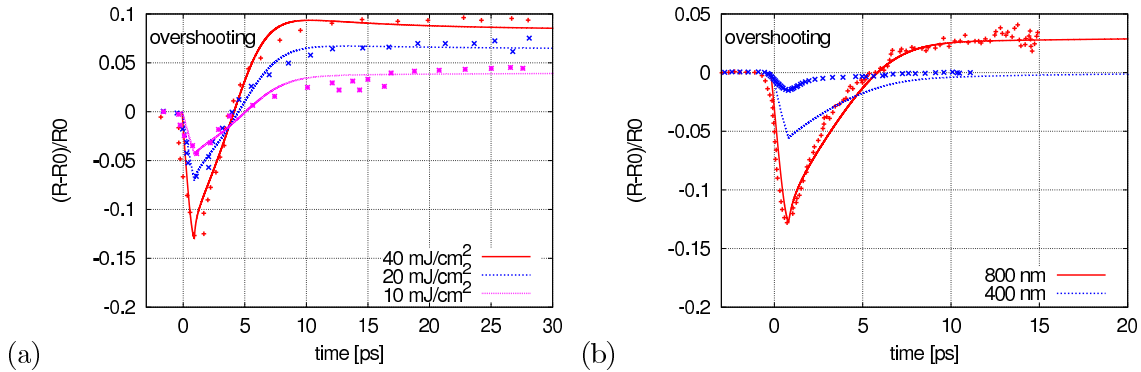}
\caption{(Color online) Relative change of transient reflectivity in GaAs as a function of time measured in: (a) the experiment by Krupin et al. \cite{krupin} at FEL photon energy of $800$ eV and optical probe of $800$ nm, (b) the experiment by Gahl et al. \cite{gahl} at FEL photon energy of $40$ eV and optical probes of $800$ nm and $400$ nm. Experimental values (points) and theory predictions (lines) are compared for: (a)  various FEL fluences: $F=10, 20,$ and $40$ mJ/cm$^2$,  and (b) various wavelengths of probe pulse. The theory results in (b) are obtained for a FEL pulse fluence of $F=4.1$ mJ/cm$^2$.} 
\label{reflectivity}
\end{figure*}

Rate equations describe the changes of electron-hole density due to ionization and recombination processes. Heat capacities of the free-electron gas  from Ref.\ \cite{ashcroft} and of  the lattice from Ref.\ \cite{ioffe} are used in the temperature equation, describing the exchange of thermal energy between free electrons and lattice. Fitted (i.e., iteratively adjusted) parameters are: (i) the free-electron temperature at the minimum of $\Delta R/R$ curve, $T_{e}^{init}$, and (ii) the thermalization time, $\tau_{el-latt}$.  Plots (Fig. \ref{reflectivity}) show the theoretical calculations (solid lines) compared to the experimental data (points).

The  ultrafast drop and rise of optical reflectivity is due to the production of electron-hole pairs in the irradiated semiconductors as described in  Ref.\ \cite{harmand}. The created free carriers contribute to the optical properties of the material. Their contribution can be described by the Drude model. 

The theoretical calculations shown in Fig.\ \ref{reflectivity}a were performed with experimentally measured fluence values. The Drude model could not be applied at the experimentally measured fluence value by Gahl et al. that lead to an overcritical electron density. However, at the pulse fluences close and larger than the one leading to the critical electron density, the minimum of $\Delta R/R$ curve does not change (Fig.\ 3b in \cite{gahl}). We then obtained results presented in Fig.\ \ref{reflectivity}b, using a fluence, corresponding to a subcritical electron density, at which the Drude model is still applicable. The predictions for $400$ nm are obtained for the same FEL pulse parameters. Note that they cannot be directly compared to the experimental curve from Fig. 2b in \cite{gahl}: according to Fig. 3a therein, it corresponds to a different set of  FEL pump  parameters. 

At a $400$ nm probe-pulse  the interband absorption coefficient is about $100$ times larger than at $800$ nm  \cite{sturge,ioffe}. The optical coefficients are sensitive to the value of absorption coefficient which, according to various estimations, may differ by a factor of  $2-3$ \cite{sturge,ioffe}. This strongly influences the accuracy of theory predictions (Fig.\ \ref{reflectivity}). Reflectivity overshooting does not show up for $400$ nm as the large absorption coefficient is then not sensitive to the small change caused by energy shift due to the band-gap shrinking ($\alpha \sim \sqrt{E_{phot} -E_{gap}}/E_{phot} $).

As we have discussed above, the characteristic time of the overshooting directly follows the timescale of the band-gap shrinking. The experimental data on reflectivity change can then be used to determine the electron-lattice thermalization time. Experimental studies of electron-phonon coupling and electron-lattice thermalization time have been extensively performed in metals \cite{hostetler, delfatti2,hohlfeld, lisowski, faure,voisin}, semiconductors \cite{delfatti,delfatti1,dean}, superconductors \cite{kardakova}, and warm-dense matter \cite{dorchies_PRL,zastrau,sokolowski}. Theoretical studies are frequently performed within the framework of two-temperature model \cite{anisimov,lin,jiang,hohlfeld,mueller,mueller1,rotenberg}. New developments can be found in \cite{ivanov, xuhong,vorberger}. 

Here, we apply our model to the experimental data to extract information on the electron-lattice thermalization time in GaAs. In Table \ref{tab} we list the  parameters iteratively adjusted to obtain the predictions in Fig. \ref{reflectivity}a, i.e.,  the thermalization time $\tau_{el-latt}$ and the free electron temperature at the minimum of $\Delta R/R$ curve $T_{e}^{init}$.
To clarify, our numerical scheme calculates transient reflectivity, adjusting iteratively: (i) the thermalization time, and (ii) the free electron temperature at the minimum of $\Delta R/R$ curve, so as to obtain the best agreement of our prediction with the experimental curve for times after the $\Delta R/R$ minimum. The average relative distance,  $D_{rel}$, between the model predictions and experimental data is the corresponding error measure. It is defined as $D_{rel}={1 \over N}\,\sum_{i=1}^N\,|x_{i,predicted}/x_{i,exp}-1|)$, where $x=\Delta R/R$ and $N$ is the number of the available experimental points. The values of thermalization time and electron temperature presented in Table 1 were obtained for the smallest $D_{rel}$ achieved which was $\sim 20$ \% for all presented parameter values. 
These values lay within the range reported in Ref.\ \cite{delfatti}. The product of collision time and photon frequency is $\omega \cdot \tau_{e} \sim$ 1 for $800$ nm   and $\omega \cdot \tau_{e} \sim$ 2 for $400$ nm. Note the increase of $T_{e}^{init}$ with fluence, and the corresponding decrease of the  thermalization time with the electron temperature \cite{delfatti}. 

\begin{table}[h]
\begin{tabular}{|c|c|c|}
\hline
$F [mJ/cm^2]\,\,\,$  & $\tau_{el-latt} [ps]\,\,\,$ & $T_{e}^{init} [eV]\,\,\,$  \\
\hline
         40 & 2.0 & 2.8   \\
\hline
         20 & 2.5 & 2.2  \\
\hline
         10 & 3.0 & 1.6  \\
\hline
\end{tabular}
\caption{Parameters used to obtain predictions in Fig.\  \ref{reflectivity}a: thermalization time  ($\tau_{el-latt}$) and the free electron temperature at the minimum of $\Delta R/R$ curve  ($T_{e}^{init}$).} 
\label{tab}
\end{table}
The values  of  $\tau_{el-latt}$ and   $T_{e}^{init}$ for  Fig. \  \ref{reflectivity}b (experiment by Gahl et al.) at the pulse fluence of  $F = 4.1$  mJ/cm$^2$ are $2.8$ ps  and   $1.6$ eV, respectively,  for both $800$ nm and $400$ nm.

\section*{Discussion}

The estimated electron temperature at the end of the electron thermalization (the minimum of $\Delta R/R$  curve) is much higher ($T_{e}^{init}\sim 2$ eV) than the temperature that we would obtain assuming a full thermalization of valence and conduction bands at this time ($T_{e}^{init}=0.4$ eV). As discussed above, this results from the delayed thermalization between electron and hole systems due to their large mass difference ($m_h/m_e\sim 10$).  A crude estimation of the average kinetic energy of a free electron within the conduction band is $\langle E_{e} \rangle = \langle E_{e-h} \rangle  -E_{gap}$, where $ \langle E_{e-h} \rangle $ is the average pair creation energy \cite{ioffe}, yields a much higher average electron temperature of  $\langle T_{e}^{init} \rangle =1.82$ eV.  This is a limiting case for which we assumed that the created hole has a negligible kinetic energy. As electron-lattice thermalization starts before the electron thermalization is completed, the estimated $T_{e}^{init}$ should be interpreted as kinetic temperature which is slightly larger than  $ \langle T_{e}^{init} \rangle $. We consider the high electron temperature as an indication for a strong non-isothermality of conduction and valence-band carriers that maintains up to a few hundred femtoseconds since exposure to the FEL pulse and  even after the carriers within each band are already thermalized. To compare, in \cite{tomassi} the heavy hole thermalization time was found to be $\sim 150$ fs within 
GaAs at room temperature. In \cite{joshi} the equilibration of  electron and hole temperatures was predicted to occur within 10 ps since the exposure to an optical pulse in a photoexcited bulk semiconductor ($Al_x Ga_{1-x} As$.)

In addition to the processes discussed above, thermal diffusion of electrons from the interaction region can influence the thermalization process. During the diffusion a fraction of electrons is slowly leaving the interaction region \cite{sundaram}, carrying a fraction of  the pulse-absorbed energy outside. 
The contribution of diffusion is related  to the attenuation length of photons which defines the size of the interaction region. For $40$ eV photons, it is $0.06$  microns, for $800$ eV photons, it is $0.6$ microns in GaAs \cite{Henke}. Created electrons can then escape to the radiation-unaffected part of GaAs bulk, 'disappearing' from the interaction region. We can therefore expect that the diffusion should affect much more the data obtained with $40$ eV photons than those obtained with $800$ eV photons. However, in both cases the estimated thermalization times and initial electron temperatures are of the same order, indicating that the effect of thermal diffusion is negligible for our data analysis.

To sum up,  in this study we presented a theoretical model that follows the relaxation of  X-ray-laser excited GaAs on a few tens of picoseconds timescale. In particular, it explains the reflectivity overshooting observed in \cite{gahl,krupin} as an effect of thermal band-gap shrinking during lattice heating. The model includes interband transitions and  uses rate and two-temperature equations to follow the relaxation of FEL excited GaAs.  Model results show that the reflectivity overshooting is a signature of electron-lattice thermalization due to electron-phonon coupling and can be used to determine the electron-lattice thermalization timescale. The presented scheme is of general applicability. We expect it to inspire dedicated, quantitative  studies  of  relaxation times in X-ray excited GaAs and other narrow band-gap semiconductors for which the thermalization and recombination times fulfill the condition $\tau_{el} << \tau_{el-latt} << \tau_{rec}$.

\section*{Methods}

The applied theoretical model uses rate equations to describe the evolution of free-carrier densities within irradiated GaAs bulk as a function of time. The rate equations are coupled with the two-temperature model \cite{anisimov}, describing the electron-lattice equilibration. The Drude model is applied to follow the  transient reflectivity as a function of carrier densities. Here, this model is extended beyond the free-carrier absorption framework \cite{dresselhaus}, and it also includes the predominant contribution from interband transitions. More details on the model are given in the Supplemental Material attached at [URL will be inserted by publisher]. Results of the model are compared to the existing experimental data on transient optical properties of GaAs taken from Refs.\  \cite{gahl,krupin}. 

\section*{References}


\section*{Acknowledgments}

\noindent
The authors thank Robin Santra for helpful comments on the manuscript.

\section*{Author contributions}

B. Z. and N. M. contributed equally to this work. B. Z., N. M. and V. T. developed the theoretical model. B. Z. carried out calculations with the model. T. M. and W. W. contributed with extensive discussions on experimental data
analysis and their interpretation. B. Z. and N. M. wrote the manuscript with contributions from all other authors.

\end{document}